\documentclass[prl,twocolumn,superscriptaddress,unsortedaddress,]{revtex4}

\usepackage{graphicx}
\usepackage{dcolumn}
\usepackage{bm}

\DeclareGraphicsExtensions{.eps,.gif,.ps}

\begin{document}


\author{Andrzej Grudka}

\affiliation{Faculty of Physics, Adam Mickiewicz University, 61-614
Pozna\'{n}, Poland}

\affiliation{Institute of Theoretical Physics and Astrophysics,
University of Gda\'{n}sk, 80-952 Gda\'{n}sk, Poland}

\affiliation{National Quantum Information Centre of Gda\'{n}sk,
81-824 Sopot, Poland}

\author{Pawe{\l} Kurzy\'nski}
\email{kurzpaw@hoth.amu.edu.pl}

\affiliation{Faculty of Physics, Adam Mickiewicz University, 61-614
Pozna\'{n}, Poland}

\title{Is There Contextuality for a Single Qubit?}

\date{May 1, 2007}%


\begin{abstract}
Cabello and Nakamura have shown [A. Cabello, Phys. Rev. Lett. {\bf
90}, 190401 (2003)] that the Kochen-Specker theorem can be applied
to two-dimensional systems, if one uses the Positive Operator-Valued
Measures. We show that the contextuality in their models is not of
the Kochen-Specker type, but it is rather a result of not keeping
track of the whole system on which the measurement is performed.
This is connected to the fact that there is no one-to-one
correspondence between the POVM elements and projectors on the
extended Hilbert space and the same POVM element has to originate
from two different projectors when used in Cabello's and Nakamura's
models. Moreover, we propose a hidden-variable formulation of the
above models.
\end{abstract}

\maketitle

For a long time there has been a debate whether a qubit is a truly
quantum system \cite{van Enk}. Although it may exist in a
superposition of two orthogonal states it does not reveal the
typical quantum oddities. The Kochen-Specker (KS) \cite{KS} and
Gleason \cite{Gleason} theorems are valid only in at least
three-dimensional Hilbert space and the Bell theorem \cite{Bell1}
applies to composite systems. Finally and most importantly  there is
a hidden-variable model describing every von Neumann measurement on
a two-level quantum system \cite{Bell}. Only recently the special
versions of KS and Gleason theorems have beem presented for a single
qubit. It was done by Cabello and Nakamura \cite{Cabello} and by
Busch \cite{Busch} respectively. The authors of both papers used the
Positive Operator-Valued Measures (POVMs) to achieve their goals.
The same year Aravind proposed how to generalize the CN method to
obtain contextual POVMs for Hilbert spaces of arbitrary dimension
\cite{Aravind}.

The essence of the contextuality is that if an observable A is
measured together with an observable B with which it commutes then
it gives different outcome than if it is measured with an observable
C with which it also commutes, and this fact has severe consequences
when one wants to describe quantum mechanics within the
framework of the hidden-variable models. If an outcome of a
measurement of an observable was preassigned before the measurement,
it would have to depend on the choice of the other observables
co-measured with the observable of interest.

In the original KS theorem A, B and C are represented by projectors,
i.e. by operators with eigenvalues 0 and 1, which makes them natural
yes-no operators. According to quantum mechanics, from a set of
mutually orthogonal projectors making up a measurement exactly one
brings the value 1 and the others bring 0. The goal of KS theorem is
to show that the pre-assignment of outcomes to a group of
measurements inevitably leads to at least one measurement wrongly
assigned.

There is no contextuality in von Neumann measurements for a
two-dimensional system (a qubit). This is because the choice of the
first projector automatically defines the second one and there is no
freedom in choosing which one to measure together with the first
projector. However, this freedom is restored if one uses POVM
measurements instead of von Neumann ones. One can have as many POVM
elements as one wants. The problem is that as for von Neumann
measurements it is natural to ask whether an outcome of a
measurement of an observable is somehow encoded in the state of the
system prior to the measurement, for a POVM it is not that simple.
To perform a POVM one has to measure an observable of a composite
system --- the qubit and an auxiliary quantum system (an ancilla)
--- thus fixing the outcome of a POVM prior to the measurement
corresponds to fixing the outcome of the measurement of the
observable on an extended Hilbert space. The important thing is that
one should not assume that the outcome of a POVM is encoded in the
qubit alone.

The projection postulate states that an act of a measurement defines
the post measurement state of the system, therefore successive
measurements of the same type should bring the same outcome. It is
true in the case of the same von Neumann measurements because the
second measurement of the system reveals the same outcome as the
first one. On the other hand POVMs are not repeatable and the second
measurement may bring a different outcome than the first one. Even
worse, in order to perform the same POVM one has to reset the state
of the ancilla, which results in preparation of the new state of the
whole system. This is another example which shows that it is hard to
speak of the KS theorem held for POVMs.

In this Letter we show that the CN model is contextual because it
does not give the complete information about the measurement. More
precisely the same POVM element acting on the original qubit can
correspond to different projectors on the extended Hilbert space. We
also give an example of CN POVM, which can be described by
non-contextual hidden-variable model of the system and the ancilla.

\begin{figure}
    {\includegraphics[width=6truecm]{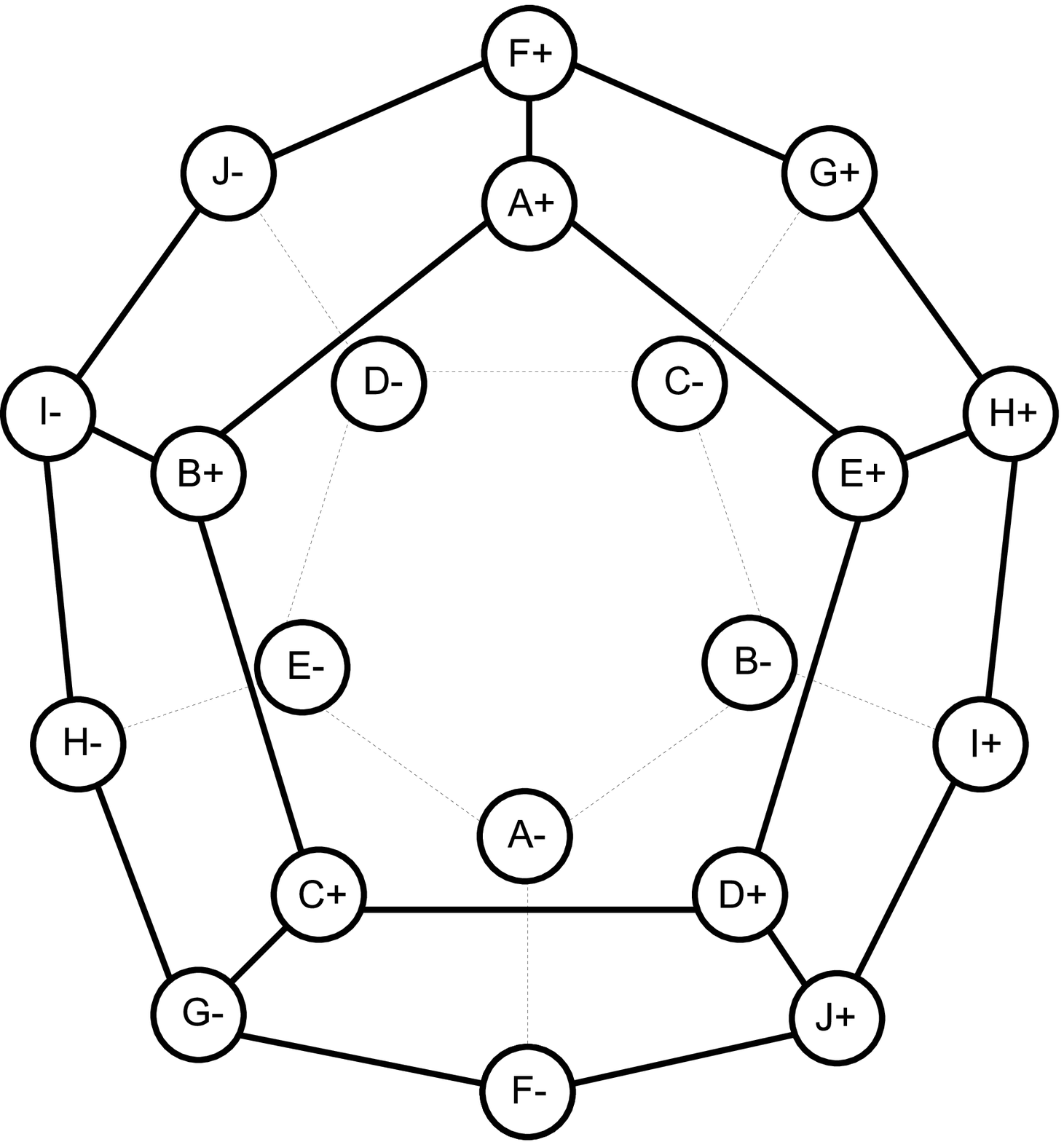}}
    {\includegraphics[width=6truecm]{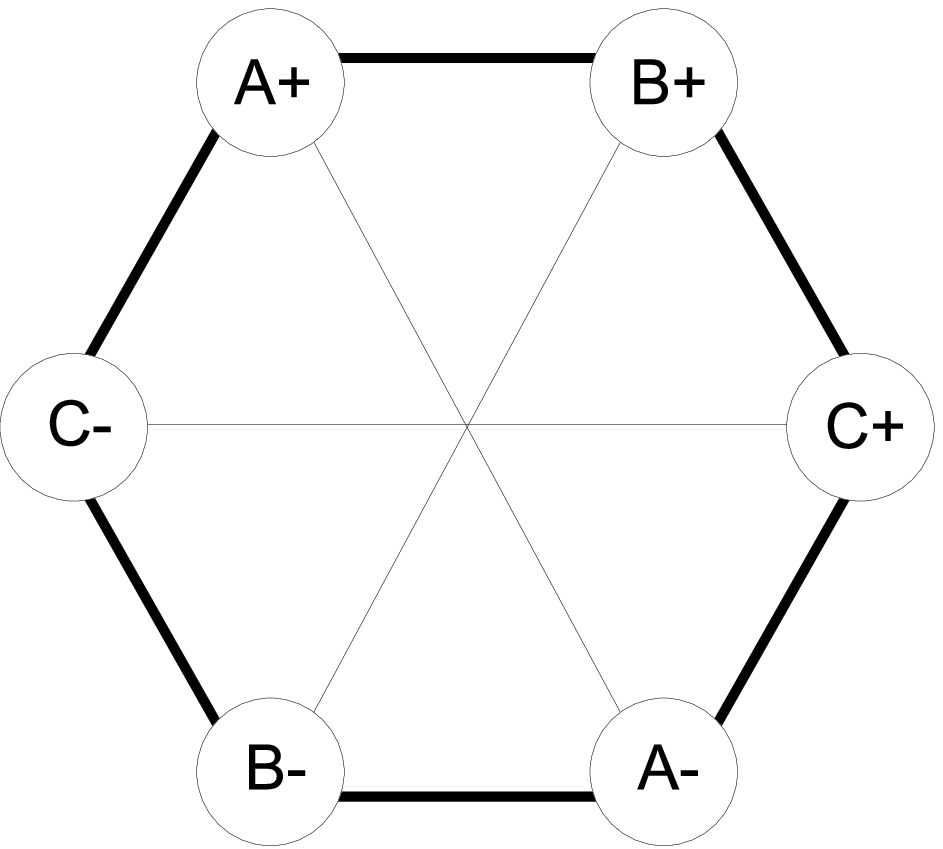}}
{ {\caption{The structure of CN POVMs. Twenty vertices of Cabello's
dodecahedron (top) and six vertices of Nakamura's hexagon (bottom).
\label{f1}}}}
\end{figure}

Let us now briefly describe the contextual set of measurements for a
qubit proposed by Cabello. His construction is based on the symmetry
of a dodecahedron enclosed in the Bloch sphere (see Fig.\ref{f1}
top). There are twenty POVM elements pointing from the center of the
dodecahedron at its vertices. These elements appear in $\pm$ pairs
pointing at antipodal vertices. Each pair appears exactly twice in
two of five eight-element POVM measurements
\begin{equation}
\begin{array}[t]{llllllllll}
\{  \varepsilon_{A_+}, & \varepsilon_{A_-}, & \varepsilon_{C_+}, &
\varepsilon_{C_-}, & \varepsilon_{I_+}, & \varepsilon_{I_-}, &
 \varepsilon_{J_+}, & \varepsilon_{J_{-}}~  \},  \\
\{  \varepsilon_{A_+}, & \varepsilon_{A_-}, & \varepsilon_{D_+}, &
\varepsilon_{D_-}, & \varepsilon_{G_+}, & \varepsilon_{G_-},
 & \varepsilon_{H_+}, & \varepsilon_{H_-}  \},   \\
\{  \varepsilon_{B_+}, & \varepsilon_{B_-}, & \varepsilon_{D_+}, &
\varepsilon_{D_-}, & \varepsilon_{F_+}, & \varepsilon_{F_-},
 & \varepsilon_{J_+}, & \varepsilon_{J_{-}}~  \},   \\
\{  \varepsilon_{B_+}, & \varepsilon_{B_-}, & \varepsilon_{E_+}, &
\varepsilon_{E_-}, & \varepsilon_{H_+}, & \varepsilon_{H_-}, &
 \varepsilon_{I_+}, & \varepsilon_{I_{-}}~  \},   \\
\{  \varepsilon_{C_+}, & \varepsilon_{C_-}, & \varepsilon_{E_+}, &
\varepsilon_{E_-}, & \varepsilon_{F_+}, & \varepsilon_{F_-}, &
\varepsilon_{G_+}, & \varepsilon_{G_-}  \}.
\end{array}
\label{e1}
\end{equation}
All elements are of the form
\begin{equation}
\varepsilon_{i}=\frac{1}{4}|\psi_i\rangle\langle \psi_i|. \label{e2}
\end{equation}
The proof of KS theorem goes as follows. By assigning 1 to any two
elements, which do not occur together, we assign values to four sets
--- the remaining elements in these sets have to be 0. Now,
the fifth set is made of elements which have already been assigned
0, since each element occurs in exactly two sets. Thus one cannot
assign a value 1 to any POVM element in the fifth set. This ends the
proof.

Nakamura proposed a more economic proof using the symmetry of a
regular hexagon (see Fig.\ref{f1} bottom). Now there are only three
$\pm$ pairs grouped in three POVM measurements
\begin{equation}
\begin{array}[t]{llllllll}
& \{  \varepsilon_{A_+}, & \varepsilon_{A_-}, & \varepsilon_{B_+}, &
\varepsilon_{B_-}  \},
  \\
  &  \{  \varepsilon_{A_+}, & \varepsilon_{A_-}, & \varepsilon_{C_+}, & \varepsilon_{C_-}  \},
\\
  &  \{  \varepsilon_{B_+}, & \varepsilon_{B_-}, & \varepsilon_{C_+}, & \varepsilon_{C_-}
\},
\end{array}
\label{e3}
\end{equation}
and all elements are given by
\begin{equation}
\varepsilon_{i}=\frac{1}{2}|\psi_i\rangle\langle \psi_i|. \label{e4}
\end{equation}
By assigning 1 to any element we assign values to two sets, and
therefore we assign 0 to all other elements. Since each element
occurs exactly twice, there is always one set in which all elements
are assigned 0.

We now show that the contextuality in both models comes from
non-unique extension of POVM elements. More precisely, POVM
measurement is performed as von Neumann measurement on the extended
Hilbert space. The restriction of von Neumann projectors leads to
POVM elements. As shown above, assignment of a physical reality to a
POVM element corresponds to the assignment of a physical reality to
a projector on the extended Hilbert space. The main point in our
argument is that two different projectors can lead to the same POVM
element and thus the contextuality comes from not keeping track of
the whole system.

First, we briefly describe the relation between the POVM and von
Neumann measurements. In order to perform POVM measurements one
extends the Hilbert space by adding an ancilla and performs  von
Neumann measurement on a higher dimensional Hilbert space (see
\cite{Peres} and references therein). Then POVM elements
$\varepsilon_{i}$ are given by
\begin{equation}
\varepsilon_{i}=\text{Tr}_{A}\{(\rho_{A}\otimes {I})P_{i} \}.
\label{e5}
\end{equation}
where $\rho_{A}$ is the state of the ancilla, $I$ is the identity on
a qubit Hilbert space, $P_{i}$ is a von Neumann projector on the
whole Hilbert space and the right hand side is traced over the
ancilla. One can always assume that the qubit and the ancilla are
unentangled, because of the identity
\begin{equation}
\text{Tr}\{(U\rho U^{\dagger}) P_i\}=\text{Tr}\{\rho (U^{\dagger} P_i U)\}.
\label{e6}
\end{equation}
which corresponds to different choice of projectors. Our task is to
examine the relation between projectors $P_i's$ that generate the CN
POVM measurements. More precisely, we want to know if a particular
POVM element which is in two different sets corresponds to the same
projector. For example, if the ancilla is a qubit in the state
$|0\rangle$ and the measurement on the whole space is done in the
Bell basis, two projectors will generate the element
$\frac{1}{2}|0\rangle\langle0|$, whereas the remaining two will
generate $\frac{1}{2}|1\rangle\langle1|$. We show that in the CN
model at least one POVM element has to be extended to two different
projectors. In order to prove this we put forward the following
hypothesis: There is a one-to-one correspondence between the POVM
elements and von Neumann projectors for each element in the CN
model. We show that this hypothesis leads to contradiction.

It should be noted that the above hypothesis includes the statement
that the state of the ancilla is the same for all POVMs, because as
already shown a change in the ancilla's state is equivalent to a
change of projectors. We begin with the Nakamura's model. For every
POVM element $\varepsilon_{X_{\pm}}$ we define the corresponding
projector $P_{X_{\pm}}$, so the measurements on the extended Hilbert
space are given by
\begin{equation}
\begin{array}[t]{llllllll}
& \{P_{A_+}, & P_{A_-}, & P_{B_+}, & P_{B_-}, & P_1\},
  \\
  &  \{P_{A_+}, & P_{A_-}, & P_{C_+}, & P_{C_-}, & P_2\},
\\
  &  \{P_{B_+}, & P_{B_-}, & P_{C_+}, & P_{C_-}, & P_3\}.
\end{array}
\label{e7}
\end{equation}
We have introduced additional projectors $P_1$, $P_2$ and $P_3$ in
order to have the resolution of the identity. Because they should
not contribute to POVM they should satisfy:
\begin{equation}
\text{Tr}_{A}\{(\rho_{A}\otimes {I})P_{i} \}=0 \label{e8}
\end{equation}
for $i=1,2,3$. All projectors in each set are mutually orthogonal.
From the first two sets (\ref{e7}) we see that both $P_{A_+}$ and
$P_{A_-}$ are orthogonal to $P_{B_+}$, $P_{B_-}$, $P_{C_+}$,
$P_{C_-}$. Since the projectors in a set span the whole Hilbert
space, from the third set we have that $P_{A_+}$ and $P_{A_-}$ have
to be confined in $P_3$, but $P_3$ does not contribute to POVM (as
we defined in (\ref{e8})), thus $P_{A_+}$ and $P_{A_-}$ cannot
contribute to POVM too. However, $\text{Tr}_{A}\{(\rho_{A}\otimes
I)P_{A_{\pm}} \}=\varepsilon_{A_{\pm}}$ which is a required
contradiction.

The contradiction vanishes if for the same POVM element one uses
different projectors in different sets. For example, if
$\varepsilon_{B_{\pm}}$ in the first POVM arose from $P_{B_{\pm}}^1$
and in the second POVM from $P_{B_{\pm}}^2$ there would be no
contradiction in the Nakamura's case.  However, this change cancels
the contextuality in the sets (\ref{e7}).

A similar contradiction can be obtained for the Cabello's model. We
introduce five projectors $P_1,P_2,\dots,P_5$ which should have the
property (\ref{e8}). The measurements on the extended Hilbert space
give:
\begin{equation}
\begin{array}[t]{lllllllll}
\{P_{A_+}, & P_{A_-}, & P_{C_+}, & P_{C_-}, & P_{I_+}, & P_{I_-}, &
 P_{J_+}, & P_{J_-}, & P_1\},  \\
\{P_{A_+}, & P_{A_-}, & P_{D_+}, & P_{D_-}, & P_{G_+}, & P_{G_-},
 & P_{H_+}, & P_{H_-}, & P_2  \},   \\
\{P_{B_+}, & P_{B_-}, & P_{D_+}, & P_{D_-}, & P_{F_+}, & P_{F_-},
 & P_{J_+}, & P_{J_-}, & P_3  \},   \\
\{P_{B_+}, & P_{B_-}, & P_{E_+}, & P_{E_-}, & P_{H_+}, & P_{H_-}, &
 P_{I_+}, & P_{I_-}, & P_4  \},   \\
\{P_{C_+}, & P_{C_-}, & P_{E_+}, & P_{E_-}, & P_{F_+}, & P_{F_-}, & P_{G_+}, & P_{G_-}, & P_5 \}.
\end{array}
\label{e9}
\end{equation}
From the first two sets we see that $P_{A_+}$ and $P_{A_-}$ are orthogonal
to $P_{C_+}$, $P_{C_-}$, $P_{I_+}$, $P_{I_-}$, $P_{J_+}$, $P_{J_-}$, $P_{D_+}$, $P_{D_-}$, $P_{G_+}$,
$P_{G_-}$, $P_{H_+}$, $P_{H_-}$. From the last three sets we can conclude
that $P_{A_+}$ and $P_{A_-}$ have to be confined in each of the three
subspaces spanned by
\begin{equation}
\begin{array}{l}
P_{B_+} + P_{B_-} + P_{F_+} + P_{F_-} + P_3, \\
P_{B_+} + P_{B_-} + P_{E_+} + P_{E_-} + P_4, \\
P_{E_+} + P_{E_-} + P_{F_+} + P_{F_-} + P_5.
\end{array} \label{e10}
\end{equation}
From the last three sets we can also conclude that
\begin{equation}
\begin{array}{l}
P_{B_+} + P_{B_-} \perp P_{F_+} + P_{F_-}, \\
P_{B_+} + P_{B_-} \perp P_{E_+} + P_{E_-}, \\
P_{E_+} + P_{E_-} \perp P_{F_+} + P_{F_-},
\end{array} \label{e11}
\end{equation}
thus $P_{A_+}$ and $P_{A_-}$ have to be confined in the subspace
spanned by $P_3 + P_4 + P_5$, which brings us again to a
contradiction.

The above reasoning shows that the contextuality in CN POVMs is a
result of not keeping track of the ancilla, i.e. if one considers a
qubit and an ancilla together, the sets of projectors are not
contextual, but if one forgets what happens to the ancilla, the
projectors become contextual POVMs. One may wonder if it is the same
kind of contextuality as the one considered by KS (for different
types of contextuality see the paper by Spekkens \cite{Spekk}). The
problem is analogous to the one we present below. Consider two
yes-no questions: {\em 1) Is it raining in Poland?} {\em 2) Is it
raining in the USA?} It is obvious that the corresponding two
answers may be different, but if we erase the ends of both questions
we will have twice: {\em Is it raining?} If the answers to the
questions are assigned before the erasure, we can have the same
question with two different answers.

Let us now present an example of CN POVM measurement for which one
can construct non-contextual hidden-variable model. Eqs. (\ref{e2})
and (\ref{e4}) define POVM elements as projectors multiplied by the
same constant ($1/2$ for the Nakamura's model and $1/4$ for the
Cabello's model). Since $\varepsilon_{X_{+}}$ is orthogonal to
$\varepsilon_{X_{-}}$, without the constant the pair would give a
von Neumann measurement on a qubit. Because of this symmetry CN POVM
measurements may be implemented by performing von Neumann
measurement on the ancilla and then, depending on the outcome,
performing one of two von Neumann measurements on the qubit in the
Nakamura's case, or one of four in the Cabello's case. The outcome of
the first measurement defines which $\pm$ pair is going to be
measured next. It is enough to prepare the ancilla in the
superposition
\begin{equation}
|\varphi\rangle=\frac{1}{\sqrt{N}}\sum_{i=0}^{N-1}|i\rangle,
 \label{e12}
\end{equation}
where $\{|i\rangle\}$ is an orthonormal basis and $N=2$ for the
Nakamura's model ($N=4$ for the Cabello's model), and then to
measure it in the basis $\{ |i\rangle \}$. There is always a
hidden-variable model of a measurement in one basis (in our case $\{
|i\rangle \}$). For example, let $\Lambda \in \{0,1\}$  ($\Lambda
\in \{0,1,2,3\}$) be a discrete hidden-variable for the Nakamura's
model (the Cabello's model). If the total description
of the ancilla is given by $\Lambda$, the
state after measurement is  $|\Lambda\rangle$ with the probability
$1/2$ ($1/4$) if  $\Lambda$ is uniformly distributed. For the next
measurement which is made on the qubit, we can describe it within
the framework of the Bell model \cite{Bell}. The corresponding
projectors on the qubit Hilbert space are:
\begin{equation}
V=\frac{
I+\vec{v}\cdot\vec{\sigma}}{2},
 \label{e13}
\end{equation}
where $\vec{v}$ is the real unit vector pointing at the $V$
direction on the Bloch sphere and $\vec{\sigma}$ is the vector of
Pauli matrices. Similarly, we define the state of a qubit as
\begin{equation} \label{e14}
|n\rangle\langle n|=\frac{ I+\vec{n}\cdot\vec{\sigma}}{2},
\end{equation}
with $\vec{n}$ being a unit vector. The hidden-variable is another
unit vector $\vec{m}$ which is uniformly distributed over all
directions on the Bloch sphere. The outcome of the measurement of a
projector (\ref{e13}) is given by
\begin{eqnarray}
 \label{e15}
v_n(V) &=& 1,~if~(\vec{m}+\vec{n})\cdot\vec{v} > 0, \nonumber \\
v_n(V) &=& 0,~if~(\vec{m}+\vec{n})\cdot\vec{v} < 0.
\end{eqnarray}

Hidden-variables allow a pre-assignment of outcomes to both von
Neumann measurements. In both cases the value 1 is assigned to
exactly one element, whereas the rest of elements are assigned 0.
The two successive measurements on the ancilla and the qubit are
equivalent to one measurement on the whole system using von Neumann
projectors of the form $| i \rangle \langle i |_A \otimes V_{\pm
Q}$, thus we can multiply the values assigned to the outcomes of the
first measurement by the values assigned to the outcomes of the
second measurement. As a result we obtain a group of measurements
which is definitely non-contextual, because in every measurement
there is exactly one element assigned to the value 1.

It is instructive to show how different extensions of POVM elements
appear in the Nakamura's model. First we consider the first set of
POVM elements. The outcome $|0\rangle$ of a measurement performed on
the ancilla's state (which was prepared in equally weighted
superposition of  $|0\rangle$ and $|1\rangle$) tells us that we
should perform a von Neumann measurement on a qubit along the
$A_{\pm}$ direction, while the outcome $|1\rangle $ tells us that we
should perform  a von Neumann measurement along the $B_{\pm}$
direction. Now for the second set if we want to have the same
extension of POVM elements the outcome
 $|0\rangle$ has to lead to a von Neumann measurement along the $A_{\pm}$ direction
 and then the outcome $|1\rangle$ leads to a von Neumann measurement along
 the $C_{\pm}$. However in the third set the outcome $|1\rangle$ cannot lead
 to a von Neumann measurements in both  $B_{\pm}$ and  $C_{\pm}$ directions.
 One of them has to be conditioned on the outcome $|0\rangle$. Thus for
 example POVM elements $\varepsilon_{B_{\pm}}$ have different extensions.

We have shown that the contextuality in the Cabello-Nakamura's model
is not of the Kochen-Specker type. It is rather a result of not
keeping track of the whole system on which the measurement is
performed. We proved that in the CN model there is no one-to-one
correspondence between POVM elements and projectors on the extended
Hilbert space. It means that some POVM elements appearing in two
different POVMs have to originate from two distinct von Neumann
projectors. Therefore, there is nothing surprising in the fact that
they bring different outcomes when measured in two different POVMs.
Moreover, POVMs are probabilistic measurements in the sense that
subsequent measurements of the same type may bring different
outcomes. The basic assumption of all hidden-variable models is that
an outcome of every measurement is pre-encoded in the state of the
system, and since in our case the whole system is not only the
qubit, it is incorrect to assume that the outcome of the POVM is
encoded only in the qubit's state. One should rather speak of
preassigning an outcome of a von Neumann measurement to the whole
system --- a qubit and an ancilla (which is not considered in the CN
model).

It would be interesting to find contextual POVMs for a qubit with a
one-to-one correspondence between their elements and von Neumann
projectors or to show that every contextual POVM on a qubit cannot
be obtained unless we give up one-to-one correspondence. We
conjecture that if the corresponding von Neumann projectors form a
contextual group of measurements one may obtain a contextual
one-to-one POVMs.

We are very grateful to Micha{\l} Horodecki, Pawe{\l} Horodocki,
Ryszard Horodecki, Micha{\l} Kurzy\'{n}ski, Marco Piani and Maria
Spychalska for useful comments and talks on the subject and the
manuscript. We would like to thank our Referees for comments and
suggestions. A. G. was supported by the State Committee for
Scientific Research Grant No. 1 P03B 014 30 and by the European
Commission through the Integrated Project FET/QIPC SCALA. P. K. was
supported by the State Committee for Scientific Research Grant No.
N202 008 31/0183. P.K. would also like to thank Tomasz {\L}uczak for
support from subsidium MISTRZ (Foundation for Polish Science).

\end{document}